# DESP-C++:
# A Discrete-Event Simulation Package for C++

Jérôme Darmont

Laboratoire d'Informatique (LIMOS)
Université Blaise Pascal – Clermont-Ferrand II
Complexe Scientifique des Cézeaux
63177 Aubière Cedex
FRANCE

E-mail: jerome.darmont@libd2.univ-bpclermont.fr

**Summary:** DESP-C++ is a C++ discrete-event random simulation engine that has been designed to be fast, very easy to use and expand, and valid. DESP-C++ is based on the resource view. Its complete architecture is presented in detail, as well as a short "user manual". The validity of DESP-C++ is demonstrated by the simulation of three significant models. In each case, the simulation results obtained with DESP-C++ match those obtained with a validated simulation software: QNAP2. The versatility of DESP-C++ is also illustrated this way, since the modelled systems are very different from each other: a simple production system, the dining philosopher classical deadlock problem, and a complex object-oriented database management system.

<u>Keywords</u>: C++ simulation package, Discrete-event simulation, Resource view, Validation

# Introduction

Many tools are nowadays available when one needs to perform random simulation. Many general simulation languages, like SIMULA [1], GPSS II [2], SLAM II [3], SIMAN [4] or QNAP2 [5] have been developed since the mid-sixties. They propose numerous functionalities and are considered valid. However, they all require a fair investment in time just to learn how to use them. When this process is complete, though, designing a simulation model is much easier than writing one from scratch. Nevertheless, dedicated simulators still remain useful when good performances are required. Furthermore, most of the general simulation languages do not allow a full object-oriented approach. Several object-oriented simulation languages and environments have been designed in the last decade, such as MODSIM II [6] and SIMPLE++ [7] that are based on C++, or Silk [8] and SimJava [9] that are based on Java. Yet, these environments still require a substantial learning time. That is why simpler simulation tools appeared in parallel as C++ or Java packages. DESP-C++ is one of them.

DESP-C++ stands for *Discrete-Event Simulation Package*. Its design originates in the modelling of object-oriented database management systems. Initially, a simulation model baptized VOODB (*Virtual Object Oriented Database*) [10] had been implemented with the QNAP2 (*Queuing Network Analysis Package* $2^{nd}$ generation) simulation software, which is a validated and reliable tool featuring a simple language close to Pascal. However, this simulation language is interpreted and our model's executions were far too slow for the intensive simulation experiments planned, especially with the object-oriented features used to extend QNAP2 [11]. Hence aroused the need for a faster simulation environment.

In short, we needed a fast, cheap, and reasonably simple object-oriented simulation language. C++ [12] obviously qualified, provided a simulation engine was coded. For simplicity's sake, we decided to implement a discrete-event random simulation kernel. It facilitated the adapta-tion of the QNAP2 VOODB model, since QNAP2 is also a discrete-event simulation software. DESP-C++ was born.

## Characteristics and functionalities

The motivation to build our own simulation engine came from the fact that the existing tools did not suit our needs, primarily in terms of validity and simplicity. The qualities we intended to give to DESP-C++ are the following.

- *Validity:* To provide reliable simulation results, DESP-C++ had to be validated, i.e., we had to check out if it was bug-free and behaved as expected. This goal was achieved by implementing the same models in QNAP2 and C++, and by verifying that the results were consistent (see the validation experiments section). Validity was a strong concern to us, and it was actually the reason why we did not select an existing C++ simulation package instead of building our own. For instance, in Reference 13, absolutely no validation experiment is provided for C++SIM. Reference 14 only proposes sample possible SimPack models, without any hint that these models are functionally correct nor that the simulator itself is bug-free.

- *Simplicity:* Provided certain basis in object-oriented programming, modelling, and simulation, DESP-C++ had to be very easy to use, compared to complex simulation software like SLAM II, QNAP2, or even other C++ simulation packages like SimPack or C++SIM, which all feature much more than what we actually needed.

- *Efficiency:* Our simulation experiments with QNAP2 being too slow, DESP-C++ had to be reasonably fast.

- *Portability:* Since our simulation models were likely to be used on several platforms (Sun, Silicon Graphics or IBM workstations; PCs under Linux or Windows), the code of DESP-C++ had to be portable. This is another reason why C++ was selected as the programming language for DESP-C++.

- *Compactness:* To remain simple and extensible, the simulation kernel had to be quite small and understandable. Its code is indeed less than 1,500 lines long, including a couple of utility functions.

- *Extensibility:* We wished to allow the possibility to develop anything, even complicated models, based on our simulation kernel. Hence we chose an "open" and simple structure that is easy to modify and expand.

- *Universality:* Many people are now comfortable with the C++ language. Hence, there is no need to learn a special syntax and starting creating simulation models is instantaneous.

- *Low-price:* DESP-C++ is a free software. Its source code is available at the following URL: *http://libd2.univ-bpclermont.fr/~darmont/download/desp-c++.tar.gz* .

DESP-C++ basically provides classes to manage and order simulation events, much like any classical discrete-event simulator. Discrete-event simulation can be defined this way [15, 16]: in a discrete-event simulation, the variables that need to be known *at all times* are *discrete*. They are called *state variables*. The set of values these variables can bear constitutes the system's *state space*. The state space is countable or finite. According to the definition of the state space, each *change in state* or *event* occurs in a discrete manner in time at instants $(t_i)_{i \in \mathbf{N}}$. These $(t_i)$ instants are called *event occurrence times* or *event dates*. Discrete-event simulation applies to any discrete system, i.e., which evolution in time is accomplished in a discrete manner. This definition covers a very broad range of systems. A couple of examples are provided in the validation experiments section. Hence, a discrete-event simulator was a natural choice to us.

Furthermore, two approaches exist to describe a system within a discrete-event simulator: the *transaction view* and the *resource view*. In the transaction view, an observer or designer de-scribes, in a chosen formalism, the behavior of the system by specifying, for each type of entity flow traversing the system, the path these entities follow and the successive operations they undergo. In the resource view, the observer describes the behavior of each active resource in the modelled system. The relationships linking active resources to various passive entities visiting them have to be defined. Among these passive entities are components that undergo some operations and passive resources that are used by active resources to perform their tasks. The transaction view was considered first, because it seemed natural with the VOODB simulation model that manages transactions in an object-oriented database system. However, coding the transaction view implied handling C++ threads, which use is not always easy. Handling threads would have played against the simplicity desired for DESP-C++. Furthermore, QNAP2 uses the resource view and we needed an easy adaptation of VOODB from QNAP2 to DESP-C++. Eventually, since any system may be modelled as easily with the resource view as with the transaction view (it is just a question of system representation), the resource view was favored for the sake of simplicity.

Thus, in DESP-C++, the system to be simulated is described by a queuing network constituted of a set of communicating resources. Resources are divided into two categories: *active resources* that actually perform some task and *passive resources* that do not directly participate in any treatment but are used by the active resources to perform their operations. The user's task is to instantiate the resource classes by specifying their parameters and their associated events, for active resources. *Clients* travel through the active resource network and are "served" by these resources.

For instance, consider a computer system where different processes run programs in parallel (Figure 1). Programs can be viewed as the system's clients and processes as the system's active resources. The processor, main memory, hard drive(s), etc. constitute this system's passive resources.

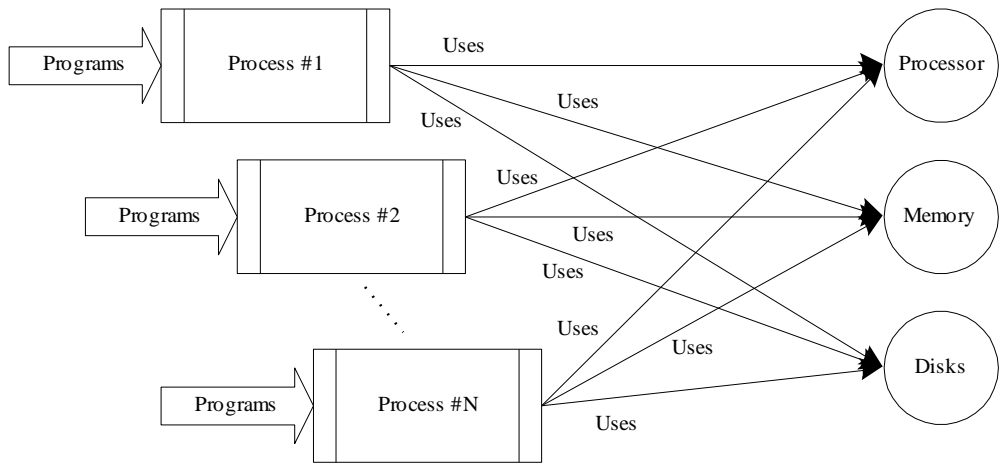

**Figure 1: Sample parallel computer system**

The behavior of a simulation model is evaluated by the mean of a set of statistics (mean values and confidence intervals). Confidence intervals are ascertained through replications of the simulation experiments using the method presented in Reference 17. By default, DESP-C++ provides the following statistics for each resource, whether passive or active:

- mean response time,
- mean waiting time for clients before being served,
- mean number of clients served,
- mean number of clients still being served,
- mean number of clients still waiting to be served.

## Architecture

The complete architecture of DESP-C++ is displayed as a UML Static Structure Diagram in Figure 2.

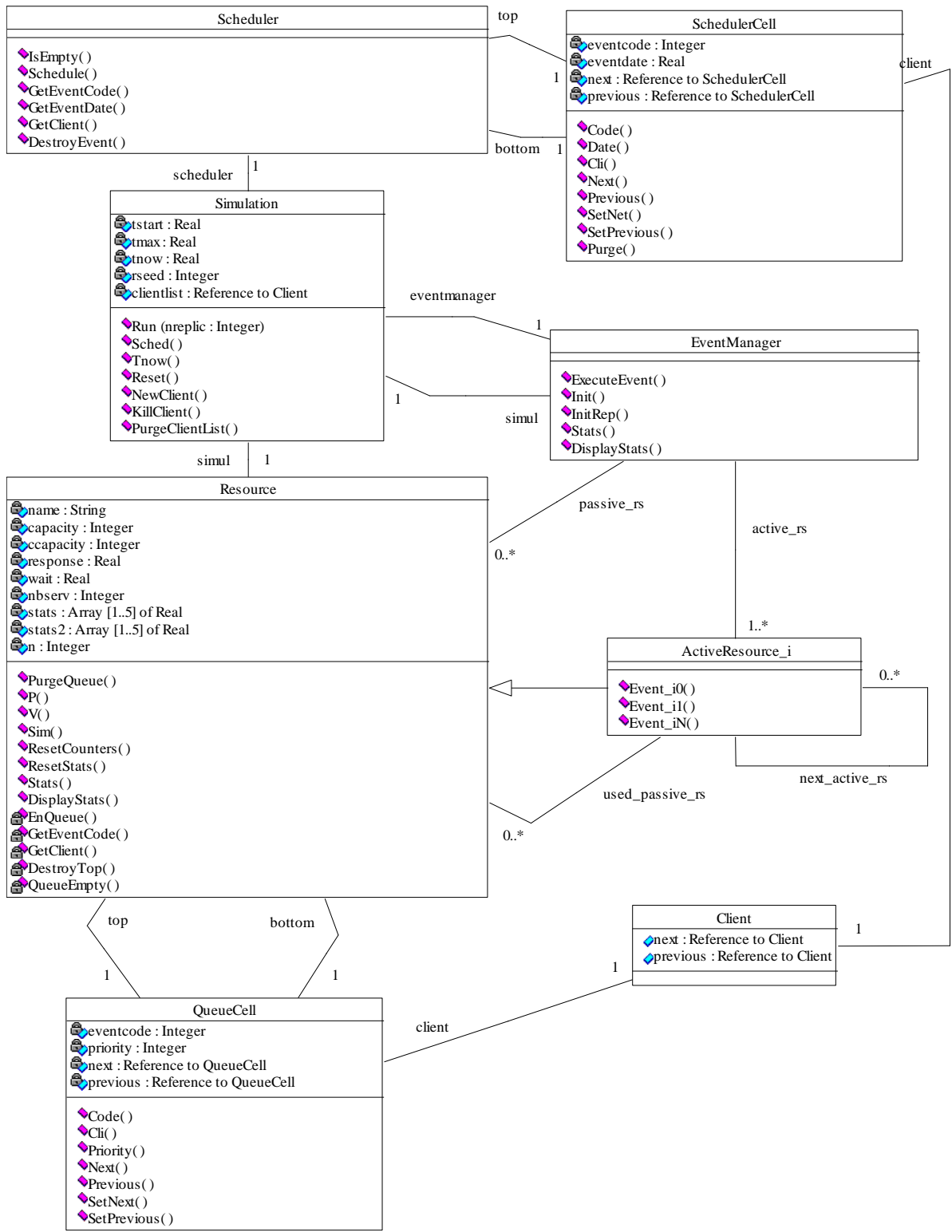

**Figure 2: DESP-C++ architecture**

DESP-C++ is organized around the *Simulation* class, whose attributes are the basic simulation control data: beginning and end of simulation times, current time, and random generator seed.

The *Simulation* class also upholds a list of references toward all the clients in the system, so that all the clients remaining in the system at the end of simulations can be destroyed and memory freed. The *Simulation* class constitutes the interface of DESP-C++. It must be instantiated in the main program. Simulation runs are activated by the *Run()* method, the number of replications being indicated in parameter.

Each simulation instance is related to a *Scheduler* object that is basically an ordered list of events to be executed, sorted by event date. Each event also has a unique code and is related to the *Client* object undergoing the event. The *Scheduler* methods deal with the event list management: event insertion, deletion and retrieval.

A *Simulation* instance is also linked to an *EventManager* object that mainly deals with the execution of events by the active resources (*ActiveResource* objects), using the passive resources (*Resource* objects), via the *ExecuteEvent()* method. The *EventManager* is also in charge of statistic initialization and computation for each resource in the system: the *Init()* method initializes the statistics for a whole simulation experiment, the *InitRep()* method does the same for one replication, the *Stats()* method computes intermediate statistics for one replication and, eventually, the *DisplayStats()* method computes and displays the final statistical results.

A *Resource* object is essentially a queue of events that are sorted by priority, each event being again associated to a *Client* object. Each *Resource* is defined by a *name* that is not necessarily unique, but ought to be, and a maximum *capacity*, i.e., the maximum number of clients it can serve concurrently. The current capacity *ccapacity* indicates how many supplemental clients may use the *Resource*. The typical *P()* and *V()* methods that are used to reserve and release the *Resource*, respectively, constitute a *Resource*'s interface along with private methods dealing with queue management (insertion, deletion, retrieval).

*DESP-C++: A Discrete-Event Simulation Package for C++*  8/28

A *Resource* also bears attributes (the *wait*, *response*, *stats[]*… counters) and methods dealing with statistics management at the individual, resource level: global initialization, initialization by replication, computation by replication, and global computation. These methods are invoked by the *EventManager* during the corresponding phases of statistics maintenance. All the active resources inherit from the *Resource* class. They just include the executions of their related events as methods in addition. Users may add extra public attributes if necessary for a particular model.

*Clients,* as mere passive entities running through the system, are just designed to be part of linked lists. However, they can be customized by users to carry any kind of information by simply adding public attributes to the *Client* class. For instance, these data can be used by active resources to perform a personalized treatment for each client.

All these classes are further organized into files and modules, as shown by Figure 3.

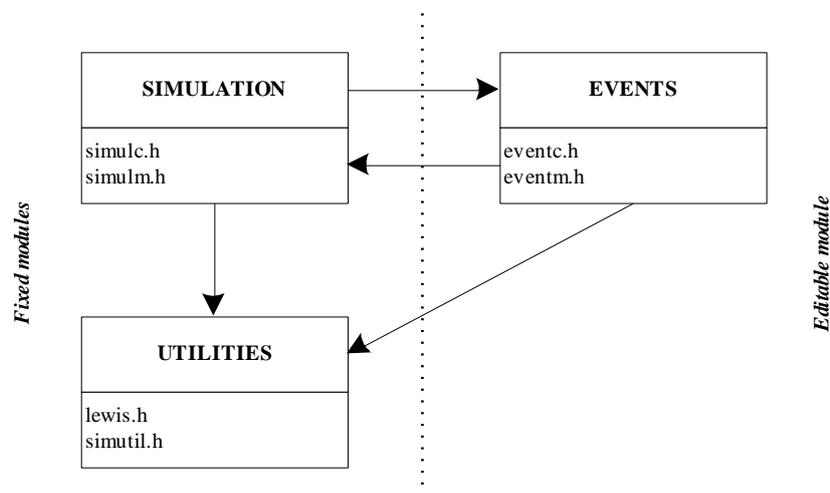

**Figure 3: DESP-C++ modules**

On the left hand of Figure 3 are the *Simulation* and *Utilities* modules, which are not normally modified by users. They contain various utilities, including an implementation of the Lewis-Payne random generator [18], which is the best pseudo-random number generator currently available thanks to its huge period, the implementation of several types of random distribution laws, and the simulation engine proper. On the right hand side stands the *Events* module, which can be modified. It deals with the definition of the system's resources, clients, and simulation events. The arrows figure how the three modules make use of each other's methods. `*c.h` files contain class definitions and `*m.h` files contain the methods' code. Other files contain utility functions.

The simulation kernel itself is very simple. Its full C++ code is presented in Figure 4 as an illustration. Basically, it functions as follows:

- global statistics are initialized;
- for each replication:
    - statistics concerning the current replication are initialized,
    - as long as the replication is not over, events are supplied by the *Scheduler* and executed by the *EventManager* — of course, events themselves do schedule other events so that the whole process iterates,
    - statistics concerning the current replication are computed,
    - all the *Client* objects remaining in the system are destroyed so that the next replication is not biased;
- global statistics are computed and displayed.

**Usage**

We strongly recommend the use of a modelling methodology like those presented References 19, 20, and 21 in order to produce correct simulation models, before any attempt to write a simulation program. Specialists in modelling and simulation at Blaise Pascal University customarily employ such a methodology, especially to model complex systems.

```cpp
// CLASS Simulation: Simulation Execution

void Simulation::Run(int nreplic) {

  int i, nextevent;
  Client *client;

  // Global initialization
  eventmanager->Init();

  // Replications loop
  for (i=1; i<=nreplic; i++) {

    // Replication initialization
    tnow=tstart;
    eventmanager->InitRep();
    client=NewClient();
    eventmanager->ExecuteEvent(0,client); // First event scheduled

    // Simulation engine
    while ((tnow<tmax) && (!scheduler->IsEmpty())) {
      nextevent=scheduler->GetEventCode();
      tnow=scheduler->GetEventDate();
      client=scheduler->GetClient();
      scheduler->DestroyEvent();
      eventmanager->ExecuteEvent(nextevent,client);
    }

    // Replication statistics computation
    eventmanager->Stats();

    // Destruction of clients still remaining in the system
    PurgeClientList();
  }
// Global results
  eventmanager->DisplayStats();
}
```

**Figure 4: Simulation kernel code**

Following a modelling methodology allows an easy and non-ambiguous specification of a given system's structure and behavior. It constitutes a guide all along the modelling process, in order to generate the most reliable models. A good use of such a modelling methodology, rather than an empirical analysis approach, induces important gains in terms of analysis time. Yet, once this modelling step is performed, translating a model in C++ is easy. Coding a discrete-event simulation model with DESP-C++ is mostly achieved by filling the *Events* module from Figure 3, i.e., by specifying the system's resources and simulation events through three steps.

Step 1: Editing the `eventc.h` file (see full code in Appendix):

- All active resources must be defined as classes inheriting from the *Resource* class (Figure 5). An active resource must "know" all the passive resources it uses (like the resource named *Passive* in Figure 5) and all the other active resources it can direct clients to.

```
// Sample active resource
class Sample_AR: public Resource {
  public:
    // Constructor
    Sample_AR(char name[STRS], int capacity, Simulation *sim, Resource *passive);
    // Events for resource Sample_AR
    void AR_Event0(Client *client);
    void AR_Event1(Client *client);
    void AR_Event2(Client *client);
    void AR_Event3(Client *client);
  private:
    Resource *Passive;
};
```

**Figure 5: Sample active resource definition**

- Pointers toward all active and passive resources must be declared as attributes of the *EventManager* class (Figure 6).

```
class EventManager {
  // Public methods (skipped)
  private:
    // Attributes
    Simulation *simul; // Pointer to Simulation object
    // Passive resources
    Resource *sample_pr;
    // Active resources
    Sample_AR *sample_ar;
};
```

**Figure 6: Resources declaration in class EventManager**

- If needed, new attributes may be added to the *Client* class (Figure 7).

```
class Client {
  public :
    // Usual attributes
    Client *next;
    Client *previous;
    // Supplementary attribute
    float operating_time;
};
```

**Figure 7: Supplementary attributes definition in class Client**

Step 2: Editing the `eventm.h` file (see full code in Appendix):

- In class *EventManager*'s constructor and destructor, respectively instantiate or destroy all active and passive resources (Figure 8).

```
// CLASS EventManager : Constructor

EventManager::EventManager(Simulation *sim) {
  simul=sim;
  // Passive resources instantiation
  sample_pr=new Resource("PR",2,simul);
  // Active resources instantiation
  sample_ar=new Sample_AR("AR",1,simul,sample_pr);
}
// CLASS EventManager : Destructor

EventManager::~EventManager() {
  // Passive resources destruction
  delete sample_pr;
  // Active resources destruction
  delete sample_ar;
}
```

**Figure 8: Instantiation and destruction of the resources**

- In class *EventManager* and method *ExecuteEvent()*, for each active resource and each event, add a line aimed at firing the event (Figure 9).

```
// CLASS EventManager : Events execution
void EventManager::ExecuteEvent(int code, Client *client) {
  switch(code) {
  case 0: sample_ar->AR_Event0(client);break; // Initial event MANDATORY!!
  case 1: sample_ar->AR_Event1(client);break;
  case 2: sample_ar->AR_Event2(client);break;
  case 3: sample_ar->AR_Event3(client);break;
  default: printf("Error: unknown event #%d at time %f\n",code,simul->Tnow());
  }
}
```

**Figure 9: Events triggering in method ExecuteEvent**

- Take all active and passive resources into account in the other methods of class *EventManager*. An example is given for method *Init()* in Figure 10.

```
void EventManager::Init() {
  // Passive resources
  sample_pr->ResetStats();
  // Active resources
  sample_ar->ResetStats();
}
```

**Figure 10: Resources' statistics initialization**

- Each active resource's constructor must be specified if it differs from the standard *Resource* constructor. Each event fired by the active resource must also be coded as a method (Figure 11).

```
// CLASS Sample_AR : Constructor

Sample_AR::Sample_AR(char name[STRS], int capacity, Simulation *sim, Resource
*passive):Resource(name, capacity, sim) {Passive=passive;}

// CLASS Sample_AR : Event #0, active resource reservation

void Sample_AR::Event0(Client *client) {
  this->P(1,client,1); // next event: #1, priority in queue: 1 }

// CLASS Sample_AR : Event #1, passive resource reservation

void Sample_AR::Event1(Client *client) {
  Resource->P(2,client,1); // next event: #2, priority in queue: 1 }

// CLASS Sample_AR : Event #2, perform operation

void Sample_AR::Event2(Client *client) {
  Sim()->Sched()->Schedule(3,Sim()->Tnow()+client->operating_time,client);
  // next event: #3, scheduled after time operating_time
}

// CLASS Sample_AR : Event #3, resources release

void Sample_AR::Event3(Client *client) {
  Resource->V();
  this->V();
  Sim()->Sched()->Schedule(0,Sim()->Tnow(),client);
  // reiterates the process now (event #0)
}
```

**Figure 11: Sample active resource methods**

Step 3: Writing a main program: this is the easy part. You just need to include the DESP-C++ modules, create a *Simulation* object and execute its *Run()* method. An example is provided in Figure 12.

```
// Sample usage program for DESP-C++

#include "simutil.h"
#include "simulc.h"
#include "eventc.h"
#include "simulm.h"
#include "eventm.h"

void main() {
  Simulation *sim = new Simulation(START_TIME, END_TIME, RANDOM_SEED);
  sim->Run(NUMBER_OF_REPLICATIONS);
}
```

**Figure 12: Sample simulation main program**

# Validation experiments

Being able to perform simulation is one thing, but obtaining reliable results is another. To achieve this goal, two conditions are mandatory:

- simulation models must be valid, i.e., they must conform to the real system they model;
- the simulator must be valid too, i.e., there must be no bug altering the results.

In order to prove that our simulation engine is adequately bug-free, we implemented the same models with QNAP2 and DESP-C++. Since QNAP2 is a valid tool, concordant results were to valid DESP-C++, making it actually "QNAP2-valid".

Though DESP-C++ is a simple tool, it is not always easy to detect and locate errors in simulation. Hence, we used testing cases that are different in terms of behavior and complexity: a simple, classical flow shop model; a little more complex model in terms of resource usage: the dining philosophers; and eventually a much more complex model: VOODB. All simulation experiments were performed on an IBM RISC 6000 workstation with 256 MB of RAM, under AIX version 4. Note that our aim here is not to validate these three simulation models but to show that our simulation engine provides unbiased results.

### *Simple model: Flow shop*

The first model concerns the flow shop production system that is presented in Figure 13. Products undergo some operations in Machine #1 for a time depending on a random exponential law (average: 10 minutes). Then, the products are transported by a mobile robot into a buffer stock ahead of Machine #2. Transport time depends on a random uniform law (values ranging from 4 to 6 minutes). The products then undergo other operations in Machine #2, for a time depending on a random exponential law (average: 12 minutes). Eventually, the products are transported outside the system by the mobile robot. Transport time still depends on a random uniform law (values ranging from 4 to 6 minutes).

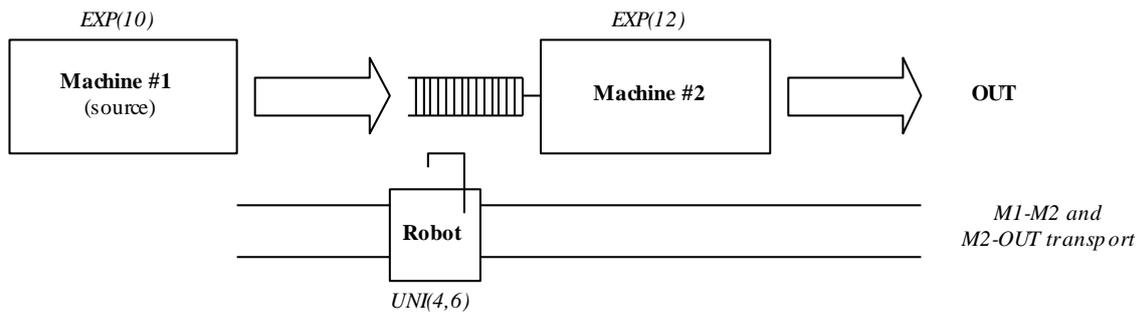

**Figure 13: Simple flow shop**

This production system is very simple. Figure 14 though illustrates the application of a modelling methodology in order to build a model of this system. This UML Activity Diagram shows the transformation process undergone by the clients (products) using the active resources Machine #1 and Machine #2, which constitute the *swimlanes* in the Activity Diagram. The passive resources do not appear on Figure 14, but they **must** also be indicated since they will be part of the simulation program code. Here, the system has only one passive resource: the robot transporting the products.

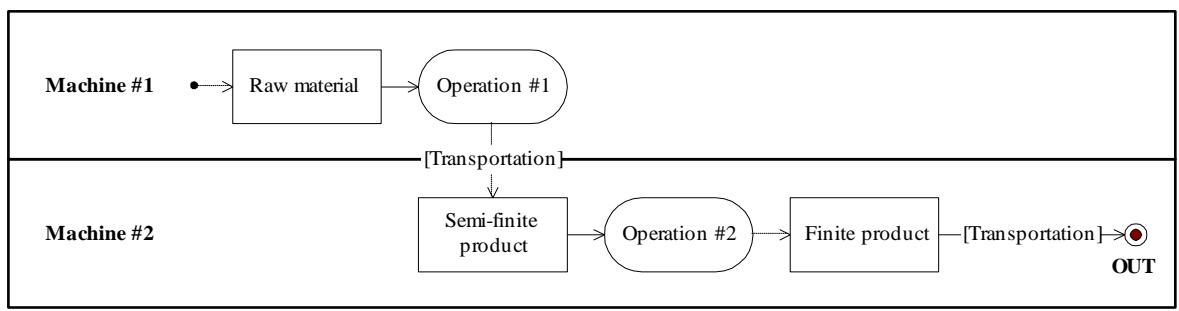

**Figure 14: Flow shop model**

To evaluate the results' conformity, the response time and the number of clients served by each resource were measured, as computed by QNAP2 and DESP-C++. The number of replications also varied from 1,000 to 15,000. The mean results obtained showed that DESP-C++ provided the same results than QNAP2 (Table 1).

|  | QNAP2 | DESP-C++ | Ratio |
|---|---|---|---|
| *Machine #1: Mean response time (min)* | 12.64 | 12.65 | 0.99 |
| *Machine #1: Average number of clients served* | 790.2 | 791.4 | 0.99 |
| *Machine #2: Mean response time (min)* | 14.73 | 14.79 | 0.99 |
| *Machine #2: Average number of clients served* | 673.6 | 672.5 | 1.00 |
| *Robot: Mean response time (min)* | 4.99 | 5.00 | 0.99 |
| *Robot: Average number of clients served* | 1463.0 | 1463.8 | 0.99 |

**Table 1: Simulation output comparison (flow shop)**

In addition, execution time for both models was measured in order to check whether the increase in performance with DESP-C++ was sufficient. On an average, DESP-C++ ran about nine times faster than QNAP2 (Figure 15).

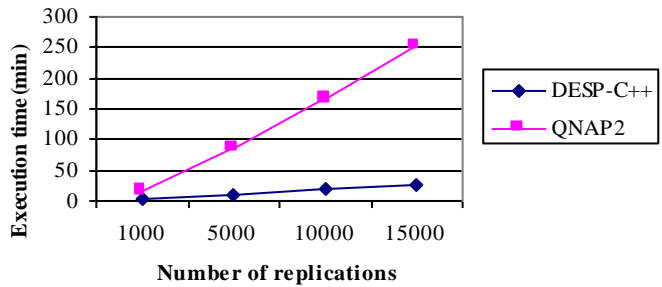

**Figure 15: Execution time comparison (flow shop)**

This study constituted a first, very encouraging validation for DESP-C++. However, we checked if the results were still as good with more elaborate models.

*Medium model: Dining philosophers*

To pursue the validation process, we considered the classical dining philosophers' problem. Four philosophers who do nothing but eat for a time depending on a random exponential law (average: 5 minutes) and think for a time depending on a random exponential law (average: 2 minutes) are seated at a table. Between each pair of philosophers is a single fork. A philosopher needs to have two forks to eat. A model for the philosopher's problem is presented in Figure 16 as a UML Activity Diagram. It describes each philosopher's behavior. Philosophers constitute the system's active resources, and forks are the passive resources.

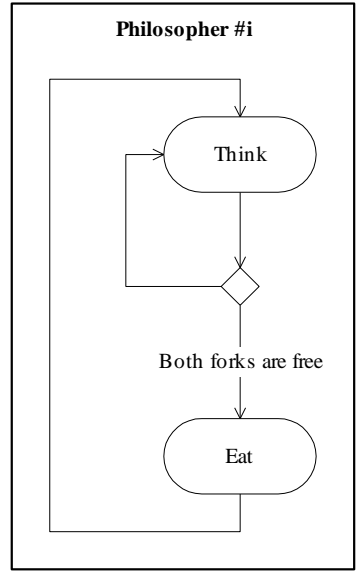

**Figure 16: Philosophers individual behavior**

We again compared the response time and the number of clients served by each resource, as computed by QNAP2 and DESP-C++, while still varying the number of replications from 1,000 to 15,000. The mean results obtained showed that DESP-C++ provided once more the same results than QNAP2 (Table 2).

|  | QNAP2 | DESP-C++ | Ratio |
|---|---|---|---|
| *Philosophers: Mean response time (min)* | 3.61 | 3.64 | 0.99 |
| *Philosophers: Average number of clients served* | 30.93 | 31.16 | 0.99 |
| *Forks: Mean response time (min)* | 5.30 | 5.32 | 0.99 |
| *Forks: Average number of clients served* | 14.58 | 14.70 | 0.99 |

**Table 2: Simulation output comparison (dining philosophers)**

Execution time was also measured for both models. On an average, DESP-C++ ran about 11 times faster than QNAP2 (Figure 17).

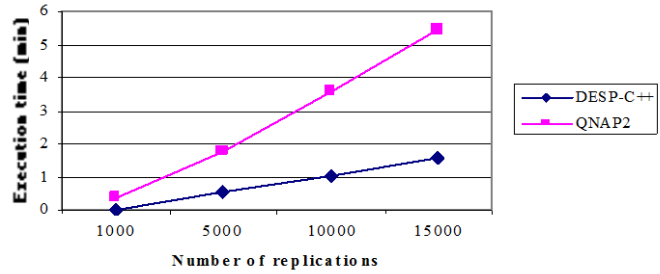

**Figure 17: Execution time comparison (dining philosophers)**

*Complex model: VOODB*

VOODB is a generic simulation model that is aimed at evaluating the performances of object-oriented database systems (OODBMSs), and more precisely, at evaluating the performances of clustering algorithms within OODBMSs. VOODB is able to model the behavior of various types of systems, especially different configurations of client-server systems.

VOODB simulates the execution of transactions within an OODB. Its workload model is constituted by the *Object Clustering Benchmark* (OCB) [22] that is a generic benchmark able to model various kinds of object-oriented databases and applications using these data. In these experiments, object bases of 50 classes and 20,000 instances were used, with four different kinds of transactions accessing the database.

Transactions are generated by *Users*, who submit them to a *Transaction Manager*. The *Transaction Manager* determines which objects need to be accessed for the current transaction and performs the necessary operations on these objects. A given object is requested by the *Transaction Manager* to an *Object Manager* that finds out which disk page contains the object. Then, it requests the page from a *Buffering Manager* that checks if the page is present in the memory buffer. If not, it requests the page from an *I/O Subsytem* that deals with physical disk accesses. After an operation on a given object is over, a *Clustering Manager* may update some usage statistics for the database. An analysis of these statistics can trigger a rec-lustering that is then performed by the *Clustering Manager*. Such a database reorganization can also be demanded externally by *Users*.

It would be too long to further describe VOODB here but a good summary is what we call the *knowledge model* for VOODB. It is presented as a UML Activity Diagram in Figure 18. This model is hierarchical and would normally be further detailed.

The knowledge model *swimlanes* figure the system's active resources. The *objects* (square boxes) represent the clients running through the system. Eventually, the *activities* (round boxes) correspond to decision rules that are invoked in the simulation events. The passive resources in VOODB do not appear here. They are the processor and main memory, the disk controller and the secondary storage, and the database itself. The clients bear several attributes, e.g., the current depth for a transaction, the OID of the next object to be accessed, etc.

The comparison between DESP-C++ and QNAP2 concerned the performances of the Texas persistent object store [23] and the DSTC clustering technique [24]. Object clustering was not included in these tests at first, to check out how everything worked out. We compared the results of 100 replications. The number of replications did not vary here since simulations with QNAP2 were already quite lengthy.

Table 3 presents the performance results obtained for a number of significant criteria. Globally, the simulation results were 97% homogeneous on an average. Computation time was about 85 times faster with DESP-C++ with this model (Table 5).

The next step was to take the DSTC clustering strategy into account within our simulation model and then to simulate the behavior of the Texas persistent object store. Performance criteria relevant to clustering were added (Table 4) and 100 replications were still performed. The results were 96% homogeneous on an average this time. This was sufficient four our needs, since simulation results are to be considered as tendencies rather than accurate values.

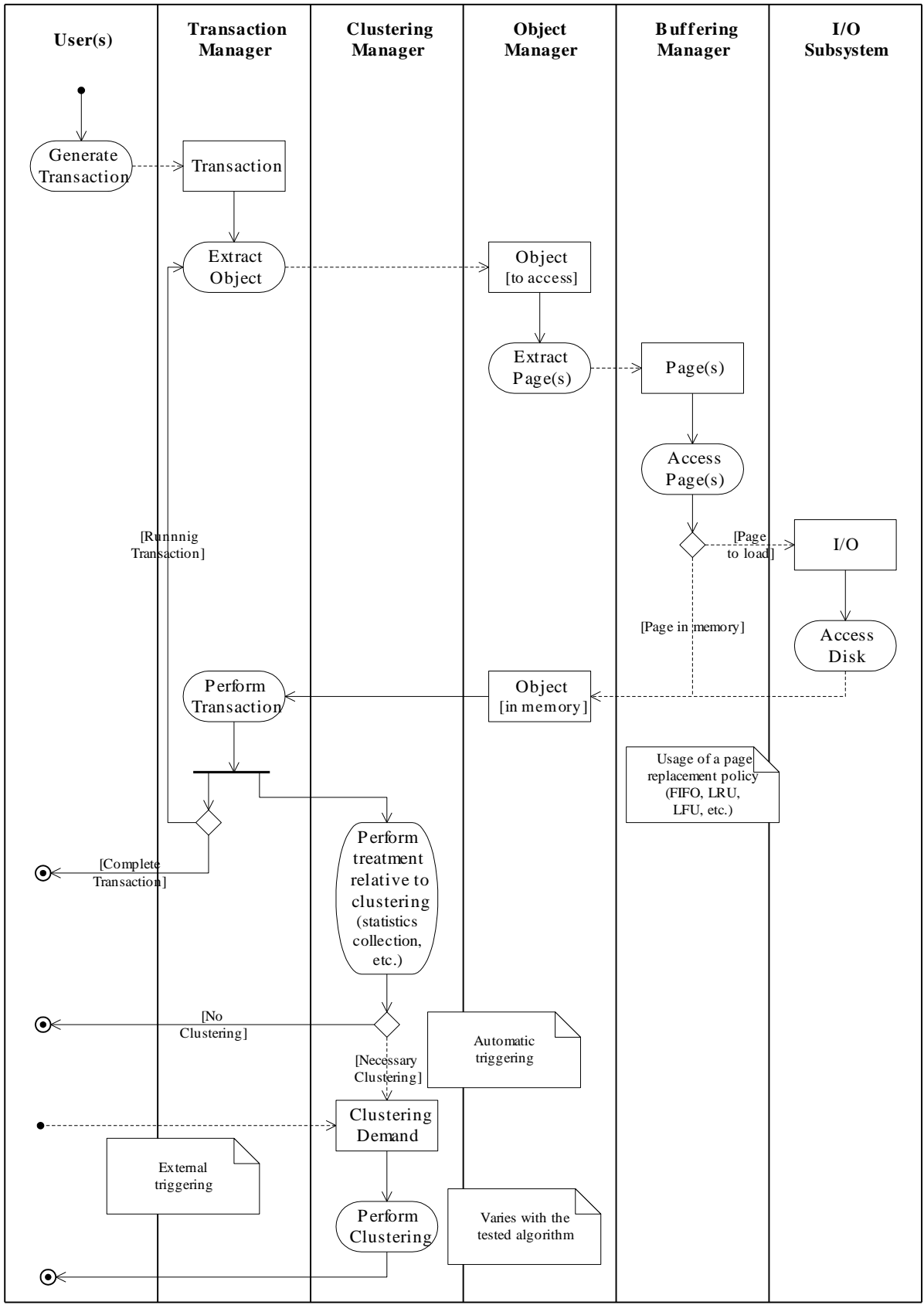

**Figure 18: VOODB knowledge model**

|  | QNAP2 | DESP-C++ | Ratio |
|---|---|---|---|
| *Mean number of transactions* | 249.4 | 250.1 | 0.99 |
| *Mean response time (s)* | 2.85 | 2.66 | 1.07 |
| *Mean number of objects accessed (per transaction)* | 64.4 | 61.5 | 1.04 |
| *Mean system throughput (transactions/s)* | 0.25 | 0.25 | 1.00 |
| *Mean number of I/Os* | 15335 | 15085 | 1.01 |
| *Mean number of disk pages used* | 2823 | 2731 | 1.03 |

**Table 3: Simulation output comparison (VOODB, no clustering)**

|  | QNAP2 | DESP-C++ | Ratio |
|---|---|---|---|
| *Mean number of transactions* | 246.0 | 250.7 | 0.98 |
| *Mean response time (s)* | 67.3 | 61.1 | 1.10 |
| *Mean number of objects accessed (per transaction)* | 2.12 | 1.86 | 1.14 |
| *Mean clustering time (s)* | 0.1 | 0.1 | 1.00 |
| *Mean system throughput (transactions/s)* | 0.24 | 0.25 | 0.98 |
| *Mean number of I/Os (transactions)* | 13073 | 12261 | 1.06 |
| *Mean number of I/Os (clustering)* | 243 | 259 | 0.94 |
| *Mean number of disk pages used* | 3066 | 3045 | 1.00 |

**Table 4: Simulation output comparison (VOODB, clustering)**

With the added complexity of clustering, the C++ model even ran almost 900 times faster than the QNAP2 model (Table 5).

|  | QNAP2 | DESP-C++ | Ratio |
|---|---|---|---|
| *No clustering* | 6,000 min. | 70 min. | 85 |
| *Clustering* | 81,000 min. | 92 min. | 880 |

**Table 5: Execution time comparison (VOODB)**

## Conclusion

We have presented in this paper an overview of the DESP-C++ discrete event random simulation engine. We discussed its main functionalities and characteristics, explained how its architecture was designed, and provided detailed usage instructions so that simulation models can be coded relatively painlessly.

We also demonstrated this tool was a valid simulation engine by comparing it to QNAP2 in terms of output. Another strong motivation was to provide a fast and easy to use simulation kernel, provided previous knowledge of the C++ language. The flexibility of DESP-C++ has been illustrated by our validation process that lead us to design three simulation models that are quite different from one another: a production system, a classical deadlock problem, and an object-oriented database management system.

Yet, there is still much room for improvement in DESP-C++. The statistical tools provided by default (replications and computation of mean values and confidence intervals) are very simple. More elaborate methods, like the regeneration or spectral methods, could achieve more reliable confidence intervals.

The mere C++ conception should also be enhanced, so that it becomes more transparent to users. A module reorganization or an implementation as a library can be envisaged. A proper graphical interface could also greatly ease the use of the package.

Eventually, some portions of code can be optimized so that simulations run even faster and data structures are more robust. Optimizing DESP-C++ was not an urge for us, but it could prove very useful. For instance, the *Scheduler* and *Resource* classes currently use basic data structures for their queues (bi-directional linked lists). More effective data structures could be used instead, like those from the LEDA [25] or STL [26] C++ libraries. STL (*Standard Template Library*) is indeed a standard C++ library since 1998 [27].

To conclude this paper, we would recommend our simulation package to people having notions of modelling and simulation, knowing the C++ language, and unwilling to learn a new language dedicated to simulation. DESP-C++ is a fair solution when one needs to rapidly and simply code a simulation model, for free.

# References

1. O.J. Dahl and K. Nygaard, 'SIMULA, an algol based simulation language', *Communications of ACM* **9(9)** (1966)

2. H. Herscovitch and T.H. Schneider, 'GPSS II – An extended general purpose simulator', *IBM System Journal* **4(3)** (1965)

3. A.A.B. Pritsker, *Introduction to Simulation and SLAM II*, Hasted Press (John Wiley and Sons), System Publishing Corporation (1986)

4. C.D. Pegden, R.E. Shanon and P.P. Sdowski, *Introduction to simulation using SIMAN*, McGraw-Hill (1990)

5. SIMULOG, *QNAP2 Reference Manual* (1995)

6. O.F. Bryan Jr., 'MODSIM II – An Object-Oriented Simulation Language for Sequential and Parallel Processors', *1989 Winter Simulation Conference*, Piscataway, NJ, 172-177 (1989)

7. AESOP GmbH, *SIMPLE++ Reference Manual* (1995)

8. K.J. Healy and R.A. Kilgore, 'Silk: A Java-based Process Simulation Language', *1997 Winter Simulation Conference, Atlanta*, GA, 475-482 (1997)

9. E.H. Page, R.L. Moose Jr. and S.P. Griffin, 'Web-Based Simulation in SimJava using Remote Method Invocation', *1997 Winter Simulation Conference*, Atlanta, GA, 68-473 (1997)

10. J. Darmont and M. Schneider, 'VOODB: A Generic Discrete-Event Random Simulation Model to Evaluate the Performances of OODBs', *$25^{th}$ International Conference on Very Large Databases (VLDB '99)*, Edinburgh, Scotland, UK, 254-265 (1999)

11. D.R.C. Hill, 'Enhancing the QNAP2 object-oriented simulation language', *Modeling and Simulation (ESM 93)*, Lyon, France, 171-175 (1993)

12. B. Stroustrup, *The C++ Programming Language*, Third Edition, Addison Wesley (1997)

13. M.C. Little and D.L. Mc Cue, *Construction and Use of a Simulation Package in C++*, Technical Report, Department of Computer Science, University of Newcastle upon Tyne, UK

14. P.A. Fishwick, *Simpack: Getting started with simulation programming in C and C++*, Technical Report #TR92-022, Computer and Information Sciences, University of Florida (1992)

15. B.P. Zeigler, *Theory of Modelling and Simulation*, John Wiley and Sons (1976)

16. A.M. Law and W.D. Kelton, *Simulation Modeling and Analysis,* $2^{nd}$ Edition, McGraw-Hill (1991)

17. J. Banks, 'Output Analysis Capabilities of Simulation Software', *Simulation* **66**(**1**), 23-30 (1996)

18. T.G. Lewis and W.H. Payne, 'Generalized feedback shift register pseudorandom number algorithm', *Journal ACM* **20(3)**, 456-468 (1973)

19. O. Balci and R.E. Nance, 'The simulation model development environment: an overview', *1992 Winter Simulation Conference*, 726-736 (1992)

20. M. Gourgand and P. Kellert, 'An object-oriented methodology for manufacturing systems modelling', *1992 Summer Computer Simulation Conference (SCSC)*, Reno, Nevada, 1123-1128 (1992)

21. P. Kellert, N. Tchernev and C. Force, 'Object-oriented methodology for FMS modelling and simulation', *Int. J. Computer Integrated Manufacturing* **10(6)**, 405-434 (1997)

22. J. Darmont et al., 'OCB: A Generic Benchmark to Evaluate the Performances of Object-Oriented Database Systems', *LNCS* **1377**, 326-340 (1998)

23. V. Singhal, S.V. Kakkad and P.R. Wilson, 'Texas: An Efficient, Portable Persistent Store', *5$^{th}$ International Workshop on Persistent Object Systems*, San Miniato, Italy (1992)

24. F. Bullat and M. Schneider, 'Dynamic Clustering in Object Database Exploiting Effective Use of Relationships Between Objects', *LNCS* **1098**, 344-365 (1996)

25. K. Mehlhorn et al., *The LEDA User Manual Version 3.7.1* (1995)

26. A. Stepanov and M. Lee, *The Standard Template Library*, Technical Report, Hewlett-Packard Company (1995)

27. Information Technology Council, X3 Secretariat, *Standard – The C++ Language*, ISO/IEC:98-14882, Washington, DC, USA (1998)

# Appendix: DESP-C++ `eventc.h` and `eventm.h` editable files

```
//
// DESP-C++ (C++ discrete-event simulation package)
// Version 1.1, February 1998
// Jerome Darmont
// LIMOS, Blaise Pascal University (Clermont-Ferrand II), France
//
// eventc.h : Definition of the Event Manager's classes
// Varies with the simulated system
//

// Active resources declaration
// Ex. class AR;

//
// CLASS EventManager
//
// Simulation events management
//

// The event manager must know all the (passive and active) resources

class EventManager {

  public:

    // Methods

    EventManager(Simulation *sim);       // Constructor
    ~EventManager();                     // Destructor
    void ExecuteEvent(int code, Client *client); // Event execution
    void Init();                         // Initialization
    void InitRep();                      // Replication initialization
    void Stats();                        // Stats computation (end of replication)
    void DisplayStats();                 // Statistics final computation & display

  private:

    // Attributes

    Simulation *simul;                   // Pointer to Simulation object

    // Passive resources
    // Ex. Resource *pr;

    // Active resources
    // Ex. AR *ar;
};

//
// CLASS Client
//
// Custom simulation entity
//

class Client {
  public :
        // Add here eventual supplementary attributes
        Client *next;
        Client *previous;
};

//
// CLASS AR
//
// Sample active resource
//
// Active resources must know all the passive resources they use and
// the "next" active resources (pointers)
```

```
//class AR: public Resource {

//   public:

//     Constructor

//     AR(char name[STRS], int capacity, Simulation *sim);

//     Events

//     void Event0(Client *client);
//     void Event1(Client *client);

//};

//
// DESP-C++ (C++ discrete-event simulation package)
// Version 1.1 g++, February 1998
// Jerome Darmont
// LIMOS, Blaise Pascal University (Clermont-Ferrand II), France
//
// eventc.h : Definition of the Event Manager methods
// Varies with the simulated system
//

//
// CLASS EventManager
//

// CLASS EventManager : Constructor

EventManager::EventManager(Simulation *sim) {

  simul=sim;

  // Passive resources instantiation
  // Ex. pr=new Resource("PR",2,simul);

  // Active resources instantiation
  // Ex. ar=new AR("AR",1,simul);
}

// CLASS EventManager : Destructor

EventManager::~EventManager() {

  // Passive resources destruction
  // Ex. delete pr;

  // Active resources destruction
  // Ex. delete ar;
}

// CLASS EventManager : Events execution

void EventManager::ExecuteEvent(int code, Client *client) {

  switch(code) {

  //case 0:  ar->Event10(client);break; // Initial event MANDATORY!!

  // Sample events

  //case 10: ar->Event10(client);break;
  //case 11: ar->Event11(client);break;

  default: printf("Error: unknown event #%d at time %f\n",code,simul->Tnow());
  }
}

// CLASS EventManager : Statistics initialization for each resource

void EventManager::Init() {

  // Passive resources
  //pr->ResetStats();
```

```
  // Active resources
  //ar->ResetStats();
}

// CLASS EventManager : Replication initialization

void EventManager::InitRep() {

  // Scheduler
  simul->Sched()->Purge();
  // Passive resources
  //pr->ResetCounters();
  //pr->PurgeQueue();
  // Active resources
  //ar->ResetCounters();
  //ar->PurgeQueue();
}

// CLASS EventManager : Statistics computation for each resource

void EventManager::Stats() {

  // Passive resources
  //pr->Stats();
  // Active resources
  //ar->Stats();
}
// CLASS EventManager : Statistics display for each resource

void EventManager::DisplayStats() {

  printf("\n*** SIMULATION STATISTICS ***\n\n");
  printf("\n*** PASSIVE RESOURCES\n");
//  pr->DisplayStats();
  printf("\n*** ACTIVE RESOURCES\n");
//  ar->DisplayStats();
}

//
// CLASS AR
//

// CLASS AR : Constructor

// AR::AR(char name[STRS], int capacity, Simulation *sim):
//    Resource(name, capacity, sim) {
// }

// CLASS AR : Event #0

// void AR::Event0(Client *client) {
//    code for event #0
// }

// ...
```